\documentclass[12pt]{iopart}
\usepackage{iopams} 
\usepackage[latin1]{inputenc}
\usepackage{graphicx}
\usepackage{url}

\def\Im{{\rm Im\mit}}

\begin{document} 
 
%\title[Giant terahertz pulling force within an evanescent field propelled by wave coupling into radiation and bound modes]%{High spatial frequency modulation of the electromagnetic energy transfer by using a graphene waveguide}
\title{Giant terahertz pulling force within an evanescent field propelled by wave coupling into radiation and bound modes} 

\author{Hern\'an Ferrari$^{1,2}$} 
\author{Carlos J. Zapata--Rodr\'iguez$^3$} 
\author{Mauro Cuevas$^{1,2}$}
\address{$^{1}$ Consejo Nacional de Investigaciones Cient\'ificas y T\'ecnicas (CONICET)}
\address{$^{2}$ Facultad de Ingeniería-LIDTUA-CIC, Universidad Austral, Mariano Acosta 1611, Pilar 1629, Buenos Aires, Argentina.}
\address{$^3$ Department of Optics and Optometry and Vision Sciences, University of Valencia, Dr. Moliner 50, Burjassot 46100, Spain.}

\ead{mcuevas@austral.edu.ar}

\begin{abstract} 
Manipulation of subwavelength objects by engineering the electromagnetic waves in the environment medium is pivotal for several particle handling techniques. In this letter, we theoretically demonstrate the possibility of engineering a compact and tunable plasmon-based terahertz tweezer using a graphene monolayer that is deposited on a high-index substrate. Under total-internal-reflection illumination, such device is shown to be capable of inducing an enhanced rotating polarizability thus enabling directional near-field coupling into the graphene plasmon mode and radiation modes in the substrate. As a result of the total momentum conservation, the net force exerted on the particle points in a direction opposite to the pushing force of the exciting evanescent field. Our results can contribute to novel realizations of photonic devices based on polarization dependent interactions between nanoparticles and electromagnetic mode fields.
\end{abstract} 



\maketitle

A well known property of light-matter interactions  is their  capacity to manipulate micro and nano particles controlling their state of motion or confine them stable in space \cite{Ashkin1,Ashkin2}. Many  configurations containing optical elements have been specially designed  for optical trapping, for controlling the rotation of trapped objects, %optical sorting,
%optically driven pumps and motors, optical delivery of
%microparticles and nanoparticles, large-scale organization of
%objects on the surface or in the space (colloidal crystals),
fusion of airborne droplets and  %ptically controlled chemical
%microreactors, 
optical chromatography \cite{zemanek1}.  % and combinations
%of optical micro-manipulation with microfluidic
%systems or optical spectroscopy (fluorescence and Raman),
%optical stretcher, and combinations of optical forces with the
%new possibilities of plasmonics and nanophotonics that
%reach into the realm of nanotechnology.

Near-field optics, that is founded in electromagnetic fields existing in the close vicinity of interfaces and whose spatial variation  is not constrained by the
diffraction limit, have found application in experiments requiring the control of the nano particle positions. Most of the literature in this field has been devoted to   configurations that use  the  total internal reflection (TIR)  phenomenon to generate evanescent waves %between a dense dielectric medium and vacuum (or liquid) 
   emerging amplified with respect to  the incident wave and taking  spatial periodicities less than those of the incident photons \cite{kawata,Lester_OL,NV1}. In those applications requiring smaller spatial variations,  the evanescent wave generated  is used to excite surface plasmons (SPs) along a metallic interface near which the 
particle or an ensamble of particles are placed  \cite{ATR1,ATR2,Marago,Juan}. %  to generate the ecanescent field interacting with the 
%particle or an ensamble of particles are placed.  

In recent years, great efforts have  been invested in  the development of structures capable to shift to the  THz  spectrum the optical phenomena taking place in the visible region. Consequently, new materials with novel optical properties in the THz range have found applications.  One of the outstanding examples is graphene, a monolayer of carbon atoms arranged
in a hexagonal lattice. Thanks to the van der Waals
force, a graphene sheet growing by CVD (chemical vapor deposition)  method can be tightly coated on the surface of a PMMA material  \cite{francisco}.   
This surface material has two attractive electromagnetic properties:  its  transparency and its capacity to supports the propagation of SPs on the electromagnetic  spectrum  from microwaves to THz.  These properties have been exploited to enhance the spontaneous emission and the electromagnetic energy transfer, as quantum nanophotonic probes \cite{mortensenAP}, as  low frequency photodetectors  \cite{mortensenPD}, PT-symmetric plasmonic waveguides \cite{OL1,soukoulis}, resonant cavities and micro antennas to enhance their sensor applications in the THz region \cite{filter}. 
In the framework of optical forces,  recently, graphene have been proposed  for THz plasmonic nanotrapping   \cite{OT0,OT1,OT2,OT22,OT3} and  for THz binding of nanoparticles \cite{FZC}. 

In the present work we show  how the two above mentioned  outstanding graphene properties:   transparency and plasmonic behavior, can be used to boost THz pulling forces. A  highlighted property of the optical pulling force is the fact that its action direction is opposed to the incident pushing force.  % forces for which the action direction is opposed to the incident pushing force.   %related with the   action direction exerted on the particle,  since the  particle acceleration results toward the source direction. 
Since its discovery, about 10 years ago, this counter intuitive  feature  has attracted the attention of the scientific  community and novel  designed light structures providing optical pulling force has been realized (see \cite{PF1} and Ref. therein). Here we propose the first transparent graphene structure demonstrating  optical pulling force in the THz spectrum. The mechanism involved in such optical phenomenon takes advantage on two properties: the asymmetric near field pattern % to the surface field 
produced by an  elliptically polarized dipole induced  on the particle beyond the critical angle of TIR, and the graphene bound mode excitation along the interface where the particle is nearly placed. Momentum transfer into radiation modes in the dielectric substrate also contributes in the resulting pulling force.   

Let us consider a single plane interface separating two   dielectric media. The semi-space $z<0$ is filled with a dense material of permittivity $\varepsilon_2=2.5$ (PMMA,  polymethyl methacrylate) and the semi-space $z>0$ is filled with a less dense medium of permittivity $\varepsilon_1$ (vacuum, $\varepsilon_1=1$). A graphene sheet is placed along the plane $z=0$ which is used to  effectively produce the optical pulling force. Since the plasmonic behavior of highly  doped graphene is the source of the  pulling force intensification, and the fact that only $p$ polarized SPs can be effectively excited in the THz range, in all the examples presented here, the illumination is accomplished by a $p$ polarized plane wave (electric field lying on the plane of incidence)  impinging on the interface $z=0$ with an angle of incidence $\phi$ from the $z$ axis,  with angular frequency $\omega$ and amplitude $E_{inc}$.

\begin{figure}%[htbp!]
    \centering
%    \subfloat[]
    {\includegraphics[width=0.95\textwidth]{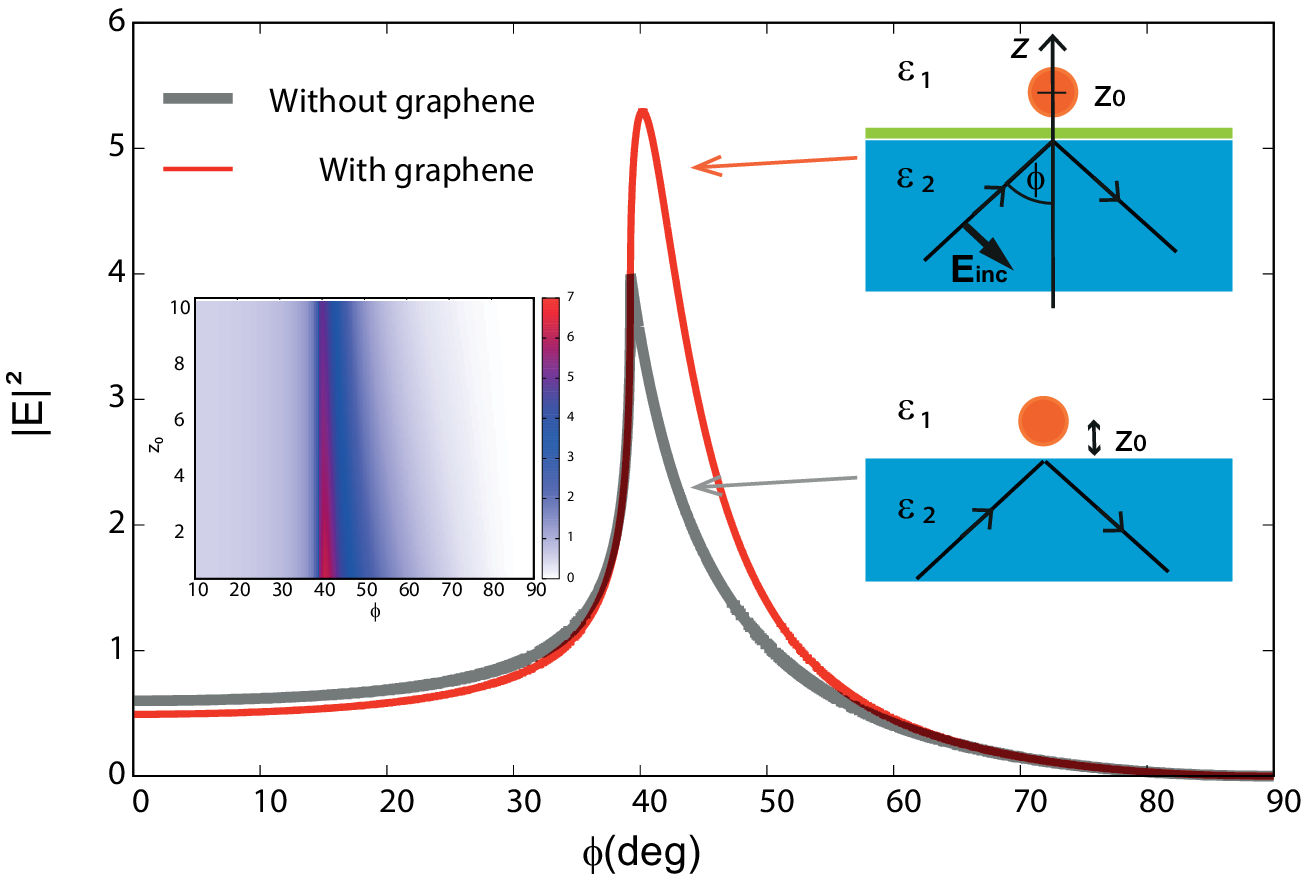}}
%    \hspace{1mm}
%    \subfloat[]{\includegraphics[width=0.45\textwidth]{OSA2.png}}
    \caption{Square modulus of the electric field (in arbitrary units)  at the particle position $z_0=1\mu$m for two cases: with  and without graphene as illustrated in the insets on the right. The permittivity of medium  1 and 2 is  $\varepsilon_1=1$ and  $\varepsilon_2=2.5$, respectively. Wavenumber of the incident field is  $\omega/c=0.05\mu$m$^{-1}$ ($\lambda \approx 126\mu$m) and the graphene parameters are   $T=295$K, $\mu_c=0.4$eV and $\gamma_c=0.1$meV. Inset: Square modulus of the electric field at the particle position $z_0$ ($1<z_0<10\mu$m) and angle of incidence $\phi$ ($0<\phi<90^\circ$). }
    \label{campos}
\end{figure}

In Figure \ref{campos} we plotted the square modulus of the electric field  at the particle position, $z_0=1\mu$m, as a function of the angle of incidence $\phi$. We observe that  for incident angles lower than the critical angle $\phi_{crit} = \arcsin (n_1/n_2)$ ($\phi_{crit}\approx 40^\circ$), the $|E|^2$ values for the cases with and without graphene almost coincide.  However, for $\phi>\phi_{crit}$, in particular for  $\phi_{crit}<\phi<60^\circ$, the square modulus of the electric field for  graphene case is approximately $1.25$ times greater than the corresponding  value  for the case without graphene. From the  inset in Figure \ref{campos} we see the dependence of the $|E|^2$ intensity as a function of $\phi$ and the particle position $z_0$ for $1<z_0<10\mu$m. %It is worth noting that  by homogeneous wave incidence the SPs along the graphene sheet do not can be excited. This is true because, at frequency   considered here, the SP propagation constant is approximately three times the value of the photon wavevector in vacuum $\omega/c$, $k_{sp} \approx 3\, \omega/c$ (see supporting information S1), a value that exceeds the photon wavevector $k_2$ in the incidence medium, $k_{2}=\sqrt{\varepsilon_2}\,\omega/c=\sqrt{2.5}\,\omega/c$. %  is grater than that corresponding to grazing incidence ($\theta=90^\circ$). 

Taking into account that the optical force is proportional to the electric field modulus (and their spatial derivative), a fact that motivated applications of enhanced evanescent fields at dielectric or metallic boundaries for increasing  the optical force \cite{Marago,NV_isa}, % on a nano particle or an ensamble of nano particles,   
%
%The high spatial localization of the evanescent field mightenable the extension of optical manipulation down to the nanometer scale
%
 one intuitively expect that the optical force  exerted on the particle in the two structures, with and without graphene  are comparable, or at most, does not differ  too much.
 %
 %1.5 times  the force value for the same system but without graphene.   
Here we reveal that, in contrast to this expectation, this is
not  the case. We find that the inclusion of graphene provides a strong interaction between the nanoparticle and the interface giving rise to a pulling force along the flat surface ($x$ direction) reaching values that are at least the same  order of magnitude than that of the pushing force produced by the enhanced evanescent field.  

In order to find the value of the force acting on the nanoparticle along the $x$ axis, we apply the Green tensor approach by assuming that the size of the particle is lower  than photon and plasmon  wavelengths, $R \ll \lambda_{sp} \approx \lambda/3$ (the photon wavevector modulus $\lambda=2\pi c/\omega$). In this framework, the time average of the total force acting on a single particle is written as \cite{NV1}, 
\begin{equation}\label{eq1}
F_x(\mathbf{r}_0) = \frac{1}{2} \mbox{Re} \sum_{j=x,y,z} p_j^* \frac{\partial}{\partial x} E_j(\mathbf{r})|_{\mathbf{r=r_0}}, 
\end{equation}
where $E_j$ is the $j$ component of the electric field, $\mathbf{p}$ is the induced electric dipole on particle at $\mathbf{r}$ position.  Taking into account that the electric  field in medium 1 can be  written as  superposition of the electric field in absence of the particle plus the field scattered by the particle, $\mathbf{E(\mathbf{r})}=\mathbf{E}^{(1)}({\mathbf{r}})+\frac{k_0^2}{\varepsilon_0}\hat{\mathbf{G}}(\mathbf{r},\mathbf{r}_0)\mathbf{p}$ ($\hat{\mathbf{G}}$ is the Green tensor of the structure), we obtain    $F_x(\mathbf{r}_0) = F_0(\mathbf{r}_0)+F_s(\mathbf{r}_0)$, where
%
%\begin{eqnarray}\label{eq3}
%F_x(\mathbf{r}_0) = %F_0(\mathbf{r}_0)+F_s(\mathbf{r}_0) 
%\end{eqnarray}
%
%where 
%
\begin{eqnarray}\label{F_0}
F_0 =\frac{1}{2} k_x \Bigg(\Im{(\hat{\alpha}_{xx})}  |\mathbf{E}^{(1)}_x|^2+\Im{(\hat{\alpha}_{zz})}  |\mathbf{E}^{(1)}_z |^2\Bigg)
\end{eqnarray}
is the $x$ component of the force due to the enhanced evanescent electric field in medium 1, and
\begin{eqnarray}\label{F_s}
F_s=-\frac{k_0^2}{\varepsilon_0} \Im{ \Bigg(\frac{\partial}{\partial x} G_{s,xz}(\mathbf{r}_0,\mathbf{r})|_{\mathbf{r}=\mathbf{r}_0}\Bigg)} \Im{(\hat{\alpha}_{xx}^*\hat{\alpha}_{zz}} \mathbf{E}^{(1)*}_x \mathbf{E}^{(1)}_z) 
\end{eqnarray}
is the $x$ component of the force due to the  interaction between the induced dipole moments and the interface, and $\hat{\alpha}$ is the particle polarizability taking into account the interaction with the interface \cite{ipor}. 

\begin{figure}[htb]
%    \centering
%    \subfloat[]
    {\includegraphics[width=0.95\textwidth]{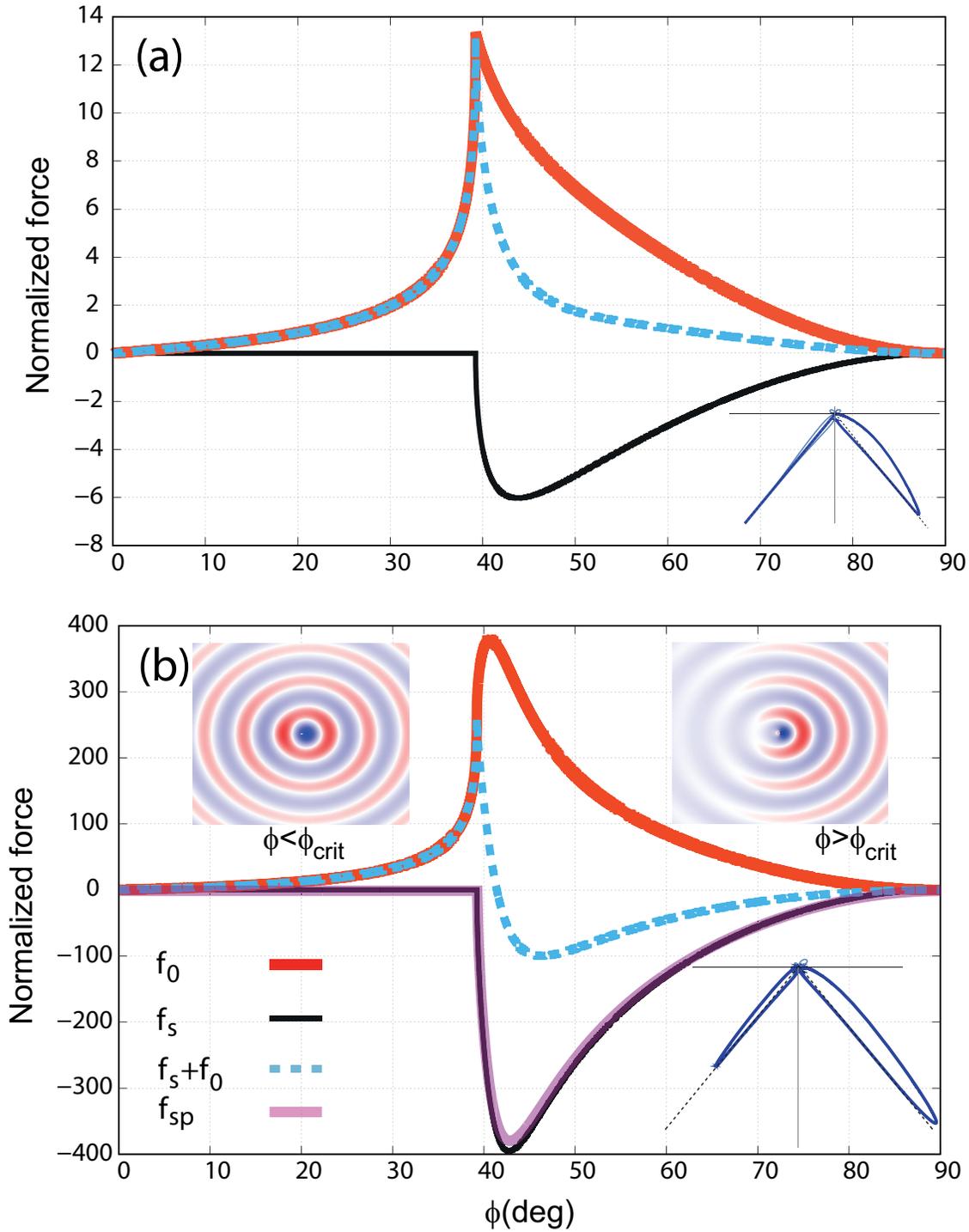}}
%    \hspace{1mm}
%    \subfloat[]{\includegraphics[width=0.45\textwidth]{OSA2.png}}
    \caption{THz force normalized with respect to the radiation force $F_{inc}$  along the $x$ axis as a function of the angle  of incidence for the case in which the interface has not graphene (a), and the case in which the interface is composed of  graphene (b). The graphene parameters are $T=295$K, $\mu_c=0.4$eV and $\gamma_c=0.1$meV. The particle (radius $R=0.045\mu$m and permittivity $\varepsilon_p=3.9$)  is placed at distance $1\mu$m from the interfece  ($z_0=1\mu$m). Inset in (b) shows the spatial distribution of the  $z$ component of the electric field along graphene for angles of incidence higher  (right) and lower (left) than the critical angle of total reflection, respectively. }
    \label{fuerzas}
\end{figure}

Figure \ref{fuerzas} shows the terahertz force   normalized with respect to the  radiation force from the incident plane wave of the same amplitude  as in Figure \ref{campos}  acting on the same particle in vacuum 
$F_{inc}=\frac{1}{2} k_0 \Im{\alpha_0} |\mathbf{E}_{inc}|^2$, \textit{i.e.}, $f_0=F_0/F_{inc}$, $f_s=F_s/F_{inc}$, $f_x=f_0+f_s$. %, $\mathbf{E}_{inc}=(E_0/\varepsilon_1)\, \hat{x} e^{i k_0 x}$.
From Figure \ref{fuerzas}a (without graphene), we observe that the curve for the force $f_0$  has the same  shape as that corresponding to $|E|^2$ in Figure \ref{campos}. %This fact can be understood from Eq. (\ref{F_0}) provided that the polarizabilities along $x$ and $z$ axis present similar values, $\hat{\alpha}_{xx} \approx \hat{\alpha}_{zz}\approx \alpha_0$ (we work far from the configurational resonances). 
We also observe 
that the total force $f_x$ is reduced with respect to $f_0$  because the $f_s$ component %, arising from the interaction with the interface 
is negative, \textit{i.e.}, $f_s$ is in opposite direction as that of $f_0$.  This fact arise from the transmitted power into medium 2 beyond the critical angle of TIR % radiatin modes  can be understood from the definition of  $G_{s,xz}$ 
(see supplementary information S2). 

In Figure \ref{fuerzas}b we plotted  the force curves when the substrate is covered  with graphene. As in the above case, the shape of the $f_0$  curve coincides with that of the enhanced evanescent field  in Figure \ref{campos}. However, the values of $f_0$ in Figure \ref{fuerzas}b  result  amplified in a factor $\approx 25-30$ with respect to those  plotted in Figure \ref{fuerzas}a. This is true because two reasons.  The first one is related with the ratio between the fields $|E|^2$  for the cases with and without graphene, which provides a factor $\approx 1.25$ (see curves in Figure \ref{campos}), and  the second one (the more important contribution)  is related to the polarizability modifications (mainly in  the imaginary part) due to self-action effect of the particle through the graphene interface. These increments are: $\Im\,\alpha_{xx}\approx 25\Im\,\alpha_0$ and $\Im\,\alpha_{zz}\approx 60\alpha_0$ (see supplementary information  S2). On the other hand, for the case without graphene,  the  increments are: $\Im\,\alpha_{xx}\approx 1.2\Im\,\alpha_0$ and $\Im\,\alpha_{zz}\approx 2.4\alpha_0$. Therefore, a rough estimation of  the ratio between the values of  $f_0$ for the cases with and without graphene gives $1.25 \times (25+60)/(1.2+2.4) \approx 30$.  

%  This fact can be understood by physical arguments as follows. Since the density of mode states is  proportional to the imaginary part of the Green tensor at the particle position,  $\Im\,G_{jj}(\mathbf{r}_0,\mathbf{r}_0)$, a fact that leads to Purcell factor increments $\approx 10^2$ \cite{NikitinSR} for graphene-particle configurations  with similar geometrical and constitutive  conditions as considered here, one should  expect the same order of increment in  the imaginary part of the polarizability $\alpha_{jj}$ (note that $\alpha_0$ is almost a real number).      
%
%These increments are: $\Im\,\alpha_{xx}\approx 25\Im\,\alpha_0$ and $\Im\,\alpha_{zz}\approx 60\alpha_0$ (see supplementary information  S3). On the other hand, for the case without graphene,  the  increments are: $\Im\,\alpha_{xx}\approx 1.2\Im\,\alpha_0$ and $\Im\,\alpha_{zz}\approx 2.4\alpha_0$. Therefore, a weak  estimation of  the ratio between the values of  $f_0$ for the cases with and without graphene gives $1.25 \times (25+60)/(1.2+2.4) \approx 30$. 

In addition,   
the force component $f_s$ %associated to the interaction with the graphene interface $f_s$ 
is two orders of magnitude higher than that corresponding to the case without graphene. 
As a result, the total force $f_x$ results negative (in the $-x$ direction) for angles of incidence $\phi>41^\circ$, \textit{i.e.},  an improved pulling force is acting on the particle.   

To find the underlying physical mechanism of the emerging pulling force, we extract the SP contribution to the $F_s$ force (\ref{F_s}) (see supporting information S1),
\begin{eqnarray}\label{F_s_sp}
F_{sp} \approx -\frac{|\alpha_0|^2 |t|^2 |E_{inc}|^2 \,  \varepsilon_2^{3/2} \sin \phi \,\kappa}{4 \varepsilon_0  \varepsilon_1^3(\varepsilon_1+\varepsilon_2)} k_{sp}^4  
\times e^{-2(k_0\kappa+k_{sp}) z_0}, 
\end{eqnarray}
where $\alpha_0$ is the particle  polarizability in vacuum, $k_{sp}$ is the propagation constant of  SPs, $\kappa=\sqrt{\varepsilon_2\sin^2\phi-\varepsilon_1}$ and $E_{inc}$ is the amplitude of the incident electric field. Note that we have used the fact that $\alpha_{xx}^*\alpha_{zz} \approx |\alpha_0|^2$ (see supplementary information S2). In Figure \ref{fuerzas} we observe that this force almost coincide with $F_s$ given by Eq. (\ref{F_s}). We attribute the small difference between $F_s$ and $F_{sp}$ to the contribution of radiative modes into medium 2 under TIR (see inset at below  in Figure \ref{fuerzas} which shows the radiation pattern for $\phi>\phi_{crit}$). %and losses on graphene to the integration defining $G_{xz}$ 
%(see supplementary information S1).
As the contrast between media 1 and 2 increases, the asymmetric radiation pattern and, consequently, the contribution of radiation modes to the pulling force becomes more evident (see supplementary information S3). 

We emphasize that the origin of the $F_{sp}$ does not arise on the SP excitation along graphene by plane wave incidence, as in attenuated total reflection devices \cite{ATR1}. %In fact, at frequency   considered here, the SP propagation constant is approximately three times the value of the photon wavevector in vacuum $\omega/c$, $k_{sp} \approx 3.2\, \omega/c$, a value that exceeds the photon wavevector $k_2$ in the incidence medium, $k_{2}=\sqrt{\varepsilon_2}\,\omega/c=\sqrt{2.5}\,\omega/c$.
Instead of this, the origin of $F_{sp}$ is due to the interaction between the elliptically dipole moment induced on the particle  by the  evanescent field and SPs on graphene. 
%
%It is worth noting that  by homogeneous wave incidence the SPs along the graphene sheet do not can be excited. This is true because, at frequency   considered here, the SP propagation constant is approximately three times the value of the photon wavevector in vacuum $\omega/c$, $k_{sp} \approx 3\, \omega/c$ (see supporting information S1), a value that exceeds the photon wavevector $k_2$ in the incidence medium, $k_{2}=\sqrt{\varepsilon_2}\,\omega/c=\sqrt{2.5}\,\omega/c$. %  is grater than that corresponding to grazing incidence ($\theta=90^\circ$). 
%
%We remark that the origin of $F_{sp}$  does not arise on the excitation of SPs by the evanescent field produced at medium 1, but due to the interaction between the nanoparticle  with an elliptically dipole moment induced  by the  evanescent field and SPs on graphene. 
%, \textit{i.e.}, the scattered by the particle field excites SPs along graphene which interact with the particle providing the pulling force $F_{sp}$. 
%For this to happen, 
Explicitly, the evanescent field  $\mathbf{E}^{(1)} \approx [ i \kappa \, \hat{x} - \sqrt{\varepsilon_2}\sin \phi \,  \hat{z} ]$,  $\phi>\phi_{crit}$, induces a dipole moment $\mathbf{p} \approx \alpha_0\,\mathbf{E}^{(1)}$ where the phase difference between the $x$ and $z$ components is equal to $-\pi/2$. %Thus,  the near field scattered by this rotating dipole excites SPs whose phase difference between the  $x$ and $z$ field components is equal to $-\pi/2$, \textit{i.e.}, SPs propagating in the $+x$ direction. 
%Thus,  the near field scattered by this rotating dipole excites SPs with an anisotropic spatial distribution,  this being more intense for $+x$ direction than for $-x$ direction (see inset in Figure \ref{fuerzas}b). 
Thus, the near field scattered by this rotating dipole excites SPs and radiation modes into the substrate with an anisotropic spatial distribution, this being more intense for $+x$ direction than for $-x$ direction (see inset in Figure \ref{fuerzas}b).
As a result of the momentum conservation, a force on the particle in the $-x$ direction appears. %Note that if $\phi<0$, the $z$ component of the evanescent field $\mathbf{E}^{(1)}$ changes its sign, and consequently,  the phase difference between the induced dipole components $x$ and $z$ 
%is $+\pi/2$ instead of $-\pi/2$. 
In virtue of the mirror symmetry respect to the $x=0$ plane, if $\phi < 0$ then the phase difference between the components $x$ and $z$ of the induced dipole turns to $+\pi / 2$, leading to a graphene SP excitation and associated force with reversed directionality.  
%Thus,  the near field scattered by this rotating dipole excites SPs whose spatial distribution is more intense for $-x$ direction, leading to a force in $+x$ direction. %   whose phase difference between the  $x$ and $z$ field components is equal to $-\pi/2$, \textit{i.e.}, SPs propagating in the $+x$ direction.
In this framework and based on the fact that a linear polarization can be decomposed as two circular polarized with opposed spins,    one would think that below TIR ($|\phi|<\phi_{crit}$)  for which the dipole moment excited on the nanoparticle has linear polarization, %two couter-propagating SPs are excited, one of them propagating in $+x$ direction and the other along $-x$ direction. As a consequence, a net null  force $F_s$ results. 
an isotropic SP spatial distribucion results (see inset in Figure \ref{fuerzas}b). % two couter-propagating SPs are excited, one of them propagating in $+x$ direction and the other along $-x$ direction. 
As a consequence, a net null $F_{sp}$ force  results. 

A pulling force assisted by SPs has also been reported in \cite{ipor} for metallic plane interfaces. However, the mechanism rising the pulling force in the system  here presented differs from that reported in \cite{ipor}. While in that work the interference of the  incident and
reflected fields on an impenetrable interface provides the necessary asymmetric excitation of SPs, in our system TIR condition is required at a transparent material  interface to reach  the asymmetric excitation of SPs  enabling the pulling force.

\begin{figure}[htbp!]
    \centering
%    \subfloat[]
%    {\includegraphics[width=0.45\textwidth]{vsLOG10z_005a10_h.png}}
    {\includegraphics[width=0.95\textwidth]{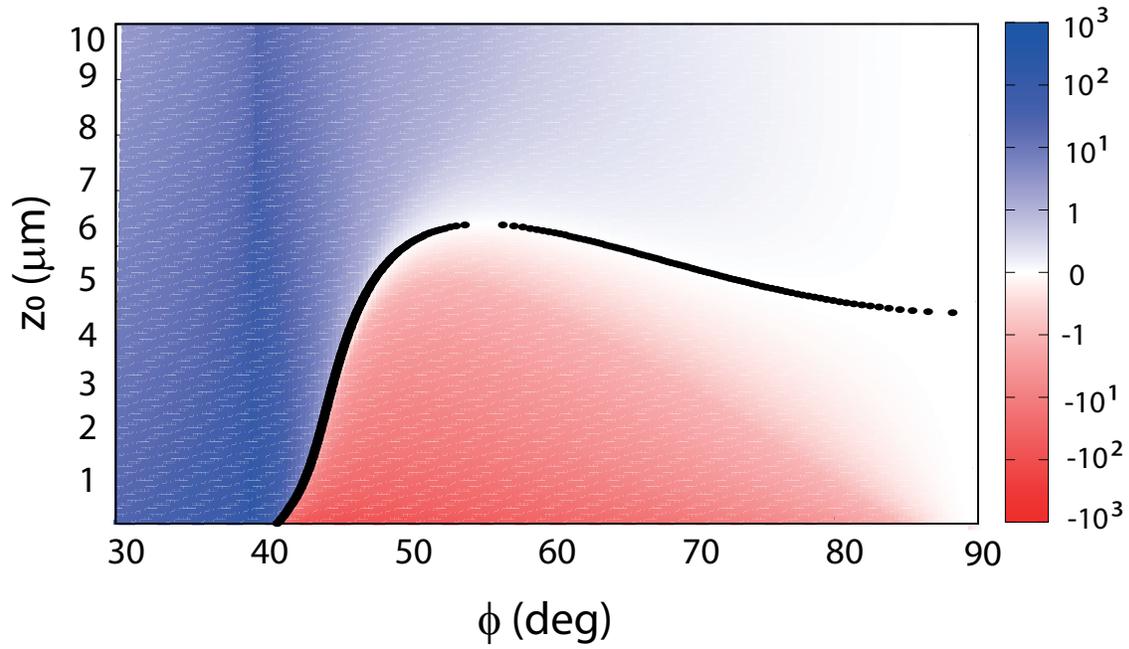}}
%figura4_OL    
%    \hspace{1mm}
%    \subfloat[]{\includegraphics[width=0.45\textwidth]{OSA2.png}}
    \caption{Normalized THz force $f_x$  as a function  of the angle of incidence $\phi$ and the $z_0$ distance.  Red, negative values, blue, positive values. Black dot corresponds to points $(\phi,z_0)$ for which the total force $f_x=0$.  All other parameters are the same as in Figure  \ref{fuerzas}b.}
    \label{fuerza_z}
\end{figure}

Figure \ref{fuerza_z} shows the normalized force $f_x$ as a function of the angle of incidence and the distance $z_0$ from the interface.  We observe two well defined  regions: one of them, the blue zone,  corresponding to points $(\phi,\,z_0)$ where the force is positive and another zone, the red zone, where the force is negative. For all angles of incidence, we note a maximum value of $z_0$  from which the  total force is positive. This occurs because the force $F_s$ is greater than $F_0$ near the graphene interface and, in addition, $F_s$ decays much faster than $F_0$. This fact can be understood 
from Eqs. (\ref{F_0}) and   (\ref{F_s_sp}) %we see that the force  $F_0$ depends of $z_0$ as 
where we see that $F_0 \approx \mbox{exp}(-2k_0 \kappa \, z_0)$ and 
$F_s \approx \mbox{exp}(-2[k_{sp}+k_0\kappa] z_0)$. Therefore, the decay distances  for $F_0$ and $F_s$ are $\delta_0 \approx 1/(2\,k_0\kappa)$ and $\delta_s \approx 1/(2 [k_{sp}+k_0 \kappa])$, respectively. As a consequence  $\delta_0>\delta_s$. 
For instance, for angles of incidence near the critical angle, \textit{i.e.}, $\sin\phi \approx \sqrt{\varepsilon_1/\varepsilon_2}$, $\delta_0 \longrightarrow \infty$, which mean that the force $F_0$ almost does not decay with $z_0$, and $\delta_s \approx 3\mu$m.   %reaching the value $\delta_0=\infty$ for the critical angle $\phi_{crit}$ and the value $\delta_0=1/(2\,\,0.05\sqrt{\varepsilon_2-\varepsilon_1})\approx 8.1\mu$m for $\phi=90^\circ$.

In conclusion, we have revealed a novel form for generating a pulling force within an evanescent field by near-field directional coupling into multiple photonic channels. In the THz spectrum, such particle handling mechanism is enabled by a simple graphene structure lying on a high-index substrate. The net pulling force is achieved through the strong field scattered by particle illumination under total-internal-reflection conditions. The magnitude of this pulling force results from the superposition of two processes: one of them provided by the interaction between the particle and the plasmon field scattered back to the particle site, and the other one driven by asymmetric excitation of forbidden propagation modes into the substrate \cite{Novotny97}. Both phenomena can be considered as a manifestation of the interaction between a rotating dipole moment and the vectorial modes of an electromagnetic structure \cite{RF1,Aiello}. Our theoretical analysis is carried out by using the Green tensor approach. We provide analytical expressions revealing the characteristics of the SP force, which is the dominant source of the pulling force, with dependence upon geometrical and constitutive parameters.

\section*{Acknowledgments} The authors acknowledge the financial supports of Universidad Austral O04-INV0 0 020 and Consejo Nacional de Investigaciones Científicas y Técnicas (CONICET ). Discussions with Prof. Francisco Ibanez  (Laboratorio de Nanoscopias y Fisicoquímica de Superficies, INIFTA, CONICET, Argentina) about CVD method and the transferring protocol involving the use of PMMA are gratefully acknowledged.

\section*{Disclosures} The authors declare no conflicts of interest.

\end{document}